# Wearable Coaxially-shielded Metamaterial for Magnetic Resonance Imaging

*Xia Zhu, Ke Wu, Stephan W. Anderson[*], and Xin Zhang[*]*

X. Zhu, K. Wu, X. Zhang

Department of Mechanical Engineering, Boston University, Boston, MA 02215, United States.

E-mail: xinz@bu.edu

S. W. Anderson

Boston University Chobanian & Avedisian School of Medicine, Boston, MA, 02118, United States.

E-mail: sande@bu.edu

X. Zhu, K. Wu, S. W. Anderson, X. Zhang

Photonics Center, Boston University, Boston, MA 02215, United States.

E-mail: xinz@bu.edu

## Abstract

Recent advancements in metamaterials have yielded the possibility of a wireless solution to improve signal-to-noise ratio (SNR) in magnetic resonance imaging (MRI). Unlike traditional closely packed local coil arrays with rigid designs and numerous components, these lightweight, cost-effective metamaterials eliminate the need for radio frequency (RF) cabling, baluns, adapters, and interfaces. However, their clinical adoption has been limited by their low sensitivity, bulky physical footprint, and limited, specific use cases. Herein, we introduce a wearable metamaterial developed using commercially available coaxial cable, designed for a 3.0 T MRI system. This metamaterial inherits the coaxially-shielded structure of its constituent coaxial cable, effectively containing the electric field within the cable, thereby mitigating the electric coupling to its loading while ensuring safer clinical adoption, lower signal loss, and resistance to frequency shifts. Weighing only 50g, the metamaterial maximizes its sensitivity by conforming to the anatomical region of interest. MRI images acquired using this metamaterial with various pulse sequences demonstrate an up to 2-fold SNR enhancement when compared to a state-of-the-art 16-channel knee coil. This work introduces a novel paradigm for constructing metamaterials in the MRI environment, paving the way for the development of next-generation wireless MRI technology.

## 1. Introduction

Magnetic resonance imaging (MRI) has emerged as a cornerstone of modern medicine, offering non-invasive, non-radioactive and high-resolution insights into the human body.[1] Its broad applications in clinical practice include disease detection, treatment planning, and therapeutic monitoring.[2,3] The underlying physics of MRI is based on the nuclear magnetic resonance (NMR) effect, where the nuclei inside human body interact with an externally applied radio frequency (RF) signal, followed by relaxation while emitting a signal that conveys information related to anatomical structure, tissue composition, pathological abnormalities, and more.[4,5] Optimally harnessing this emitted signal hinges on the development of the RF subsystem, which includes a dedicated receive coil or receive coil array[6] positioned in close proximity to the human anatomy of interest. This proximity enhances sensitivity, resulting in a robust signal-to-noise ratio (SNR).

Conventional RF receive coil arrays typically have bulky, fixed, and rigid designs, as shown in **Figure 1**a. Due to the need for compatibility with various anatomical sizes, these arrays inevitably pose challenges in patient comfort and positioning, as well as suboptimal signal sensitivity in some cases. To address these issues, the MRI receive coil design paradigm has evolved toward flexible and wearable designs. Recent fabrication techniques have introduced various flexible conductor trace approaches, such as screen-printed conductors,[7,8] liquid metal tubes,[9,10] copper braids,[11] conductive threads,[12] and conductive elastomers,[13] as viable alternatives to traditional coils. These designs offer similar SNR while improving conformability, thereby enhancing patient comfort and providing versatility for specific applications. However, similar to conventional rigid coils, these flexible designs often incorporate numerous non-magnetic electronic components, feed boards, cable traps, and adapters, adding to their overall bulkiness and expense while requiring careful handling during routine imaging procedures.

The emergence of metamaterials,[14,15] a class of artificially constructed materials composed of sub-wavelength unit cells, has yielded a new perspective on enhancing SNR using a wireless, cost-effective solution. Leveraging their ability to tailor and redistribute incident electromagnetic waves in the near field region, metamaterials have found widespread applications, including cloaking devices,[16] perfect absorbers,[17] wireless power transfer,[18,19] and high-sensitivity sensing.[20] Notably, their magnetic field enhancement capacity has enabled them to focus circularly polarized MRI signals for increased SNR. Various metamaterial configurations, such as 'conducting Swiss Rolls,'[21] capacitively-loaded ring resonators,[22,23] helical coils,[24-26] and parallel wire resonators,[27-29] have been proposed. However, their clinical adoption is hindered by their limited sensitivity[24,26] (in comparison to state-of-the-art receive coil arrays) and, in some cases, incompatibility with commonly used clinical sequences.[27,28] Furthermore, most of these proposed structures employed rigid and bulky designs,[29,30] which substantially restrict both achievable sensitivity and patient comfort. Consequently, a fundamental paradigm shift in the metamaterial design is needed for seamless clinical integration of these wireless devices.

Coaxial cables, known for their signal quality preservation over long distances, have traditionally been employed for signal transmission in various RF scenarios, including MRI. By leveraging their Faraday shielding effect from the cable's outer conductor, a series of shielded loop resonators for wireless non-radiative power transfer (WNPT) have been proposed.[31-33] These delicately designed resonators offer remarkable magnetic response by eliminating the electric dipole moment found in conventional helical coils used for WNPT.[34] Additionally, transmission line resonators crafted using coaxial cables have found applications in diagnostic MRI,[35-37] offering a reduced dielectric loss and capacitive detuning when imaging conducting samples, while conforming smoothly to the human body's contours, enhancing patient comfort. Recent advancements have seen the development of high-impedance MRI detectors array utilizing coaxial cable technology.[38-43] These detectors effectively isolate themselves electromagnetically from neighboring elements, achieving desired decoupling effects. It is worth noting that these implementations of coaxial cables have typically been hardwired to a direct RF feed and are not operated wirelessly. Nonetheless, the advantages of coaxial cables, including minimized dielectric loss, resilience against frequency detuning, and their inherent flexibility and ease of handling, position them as exceptional candidates for constructing magnetic metamaterials within the context of MRI technology.

In this study, we introduce a novel approach using off-the-shelf coaxial cables to create a magnetic metamaterial with a primarily magnetic response for wirelessly enhancing SNR (Figure 1b). This innovative design features an array of wireless, coaxially-shielded loop resonators, capable of functioning collectively and focusing RF electromagnetic signal at a sub-wavelength scale while achieving an SNR comparable to state-of-the-art receive coil arrays. We introduce an additive multi-segment diode-loaded ring resonator (DLRR) to the metamaterial, enabling passive detuning of the resonance mode to enhance compatibility with various clinical sequences. The metamaterial and the DLRR are seamlessly integrated into a thin, lightweight and wearable fabric for ease of implementation and patient comfort. The performance of the metamaterial is rigorously characterized through simulation, on-bench measurement and testing on a clinical 3.0 T MRI system.

## 2. Results

### 2.1. Design strategy of the metamaterial

Commercially available coaxial cable (9849, Alpha Wire) with a diameter of 2.54 mm was employed to construct the metamaterial. A thin, lightweight and wearable fabric (80% nylon, 20% Spandex) served as the substrate for housing the metamaterial, as illustrated in Figure 1c. Two layers of the fabric were securely bonded using a digitally embroidered pattern to generate stretchable ducts into which the metamaterial was inserted and assembled. Figure 1d provides a cross-sectional view of the metamaterial unit cell. The adopted coaxial cable consists of four layers: the inner conductor, the fluorinated ethylene propylene (FEP) insulating layer, the outer conductor and the FEP jacket. The unit cell was formed by welding the inner and outer conductor of the coaxial cable at the open slit, which was then reinforced with heat shrink tubing. The inner conductor was intentionally left open-circuited to allow for resonating current excitation. A hexagonal array comprising 7 unit cells was constructed in this manner. Along the periphery of the metamaterial array, the DLRR, constructed from five coaxial cable segments, was integrated to facilitate strong inductive coupling with the metamaterial. Figure 1e displays a cross-sectional view of the junction between two segments, where their inner conductor was welded to the outer conductor of adjacent segment. A PIN diode (MADP-000235-10720T, MACOM Technology Solutions) was inserted in each segment to bridge the inner and outer conductors. In total, five identical junctions were integrated along the DLRR.

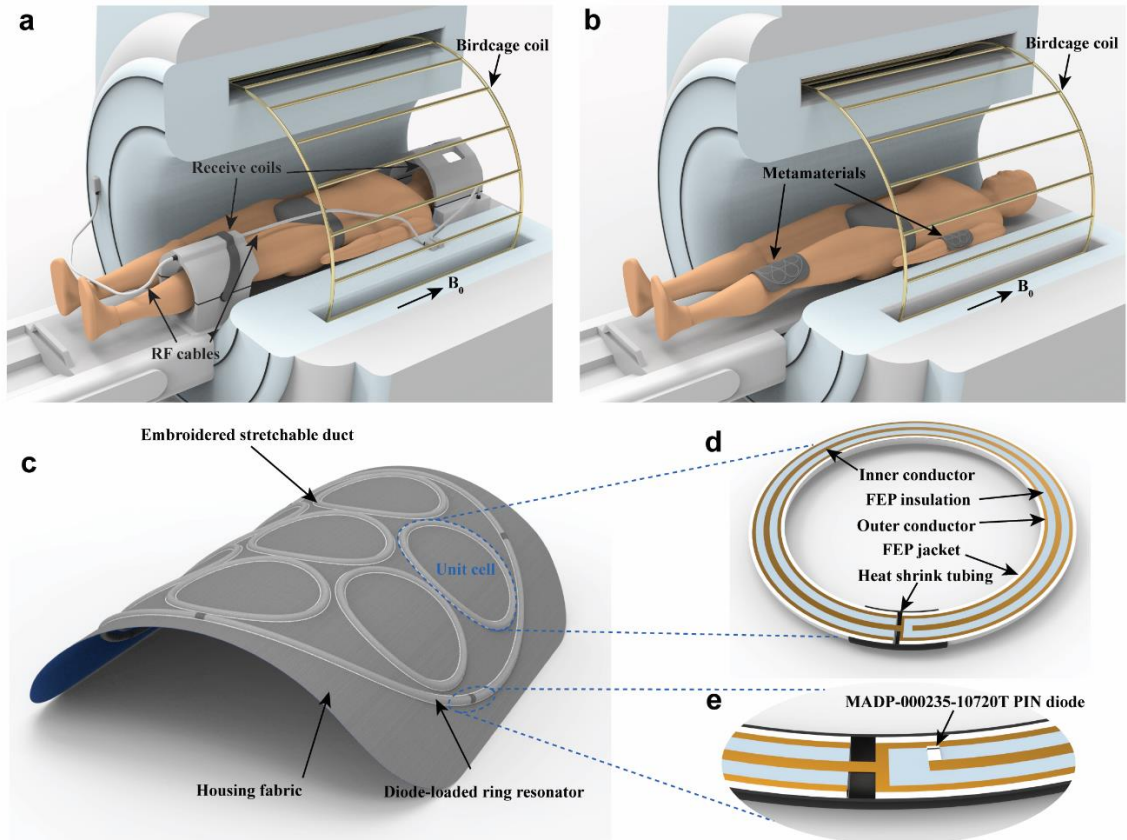

**Figure 1.** Concept and configuration of the coaxially-shielded metamaterial. a) Schematic of a conventional MRI scan in clinical practice, involving fixed and rigid local receive coil arrays, hardwired to the MRI machine via RF cabling for signal acquisition. b) Schematics of the proposed wearable, coaxially-shielded metamaterial, closely form-fitted to the patient anatomy while operating wirelessly. c) Schematics of the metamaterial, integrated into the housing fabric, being flexed into a semi-cylindrical configuration. d) Cross-sectional views of the unit cell and e) a junction in the DLRR.

The employed coaxial cable is inherently flexible, allowing moderate deformation and bending of the metamaterial, as conceptually depicted in Figure 1c. This flexibility enables the metamaterial to be form-fitting when imaging a curved anatomical region such as the human knee, ankle or wrist. In contrast to conventional fixed and rigid coil arrays, the metamaterial's form-fitting nature ensures better sensitivity by reducing the separation between the coil elements and the patient's anatomy. The metamaterial operates through wireless inductive coupling to the birdcage coil (BC), eliminating the need for excessive cabling, electronic panels, interfaces, adapters, or baluns. Consequently, the metamaterial is exceptionally lightweight, weighing only 50 g, including the housing fabric. This characteristic significantly enhances patient comfort during prolonged scanning sessions.

## 2.2. Electromagnetic characteristics of the metamaterials

The primary advantage of the coaxially-shielded metamaterial, compared to previously adopted designs,[22,23,30] lies in its predominantly magnetic response, which is due to the unique behavior of its constituent coaxial cabling described in detail below. The metamaterial incorporates no lumped elements along the cable conductor, avoiding additional losses and promoting a high quality factor. Its self-resonance is solely dictated by its inherent geometric properties. By leaving the inner conductor open-circuited at one end, a substantial structural capacitance is formed between the inner and outer conductors, enabling resonance despite the metamaterial's small electrical size relative to the RF wavelength in 3.0 T MRI (2.35 m). **Figure 2**a illustrates the cross-section of the coaxial cable. When the metamaterial is on resonance, three distinct currents are present within the structure. On the surface of the inner conductor, a current ($I_i$) flows while increasing linearly from the open end to the welding junction. Simultaneously, an inductive current of the same magnitude but opposite direction ($I_{oi}$) travels along the inner surface of the outer conductor. Additionally, a parasitic loop is formed and connects the inner conductor and the inner surface of the outer conductor through the outer surface of the latter. Due to the skin depth effect, a current ($I_{oo}$) nearly uniformly traverses the outer surface of the outer conductor. Figure 2b presents electromagnetic simulation results of the three surface current densities on a single metamaterial unit cell. These currents are then integrated along the radial direction of the cable cross-section, with the results displayed in Figure 2c. It is evident that $I_i$ and $I_{oi}$, while flowing in opposing directions, possess the same amplitudes across the entire resonator loop, effectively canceling each other with respect to the inductive magnetic field, resulting in no net contribution when observed at a distance from the metamaterial. In contrast, the current $I_{oo}$ is uniformly distributed with a magnitude equal to the maximum value of $I_i$ and $I_{oi}$, making it the sole contributor to the inductive magnetic field. The surface current densities in each segment of the DLRR follow a similar trend as the current in a metamaterial unit cell, resulting in a uniform $I_{oo}$ along the entire DLRR, as demonstrated in Figure 2d. Further details on the current distribution can be found in Figure S1 of the Supporting Information. In fact, the behavior of $I_i$ and $I_{oi}$ closely resembles the concept of the differential mode signal often discussed in coaxial cable transmission lines.[44] The differential mode carries the desired signal while being shielded from the external environment by the cable's outer conductor. On the other hand, $I_{oo}$ behaves similarly to the common mode that travels along the outer shield of the coaxial cable transmission line. In various radio frequency systems, including MRI systems, the common-mode signal is undesired and is often suppressed using baluns and cable traps.[45,46] However, in our proposed metamaterial, the intentional induction of the 'common-mode' $I_{oo}$ on the metamaterial serves to generate a secondary magnetic field, thereby achieving an enhanced SNR.

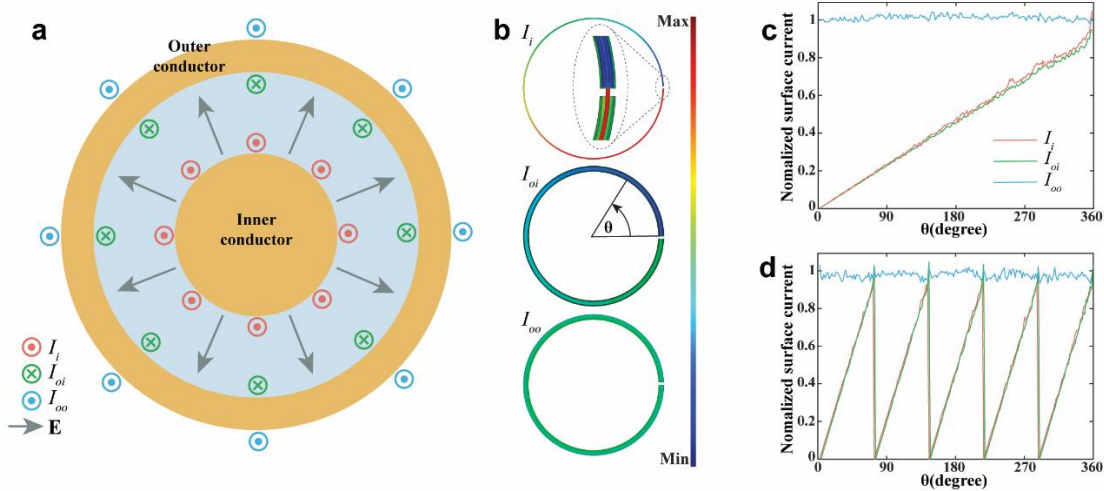

**Figure 2.** Characterization of the current inside the metamaterial. a) Current distribution within the coaxial cable of the resonating metamaterial. b) Simulated surface current density on the inner conductor ($I_i$), the inner surface of the outer conductor ($I_{oi}$) and the outer surface of the outer conductor ($I_{oo}$). c) Integrated surface currents distribution along the loop of a unit cell and d) the DLRR.

In addition to the advantages mentioned above, one of the most important characteristics of the coaxially-shielded metamaterial is its effective confinement of the electric field between the two layers of conductors, where the structural capacitance forms, as indicated by the solid arrows in Figure 2a. Typically, magnetic metamaterials may enhance the magnetic and electric field simultaneously during MRI scans.[47] While the magnified magnetic field directly impacts the SNR, the enhancement of the electric field raises safety concerns. MRI safety primarily hinges on the specific absorption ratio (SAR), which is proportionate to the square of the electric field.[48] The confinement of the electric field by the coaxial cable substantially diminishes its presence in the space around the metamaterial, where patients are exposed during MRI scans. Furthermore, the reduced electric field minimizes capacitive coupling to the sample by decreasing the induced eddy current within the sample. Therefore, our design approach makes the metamaterial both safer and less susceptible to frequency detuning induced by loading, which results in lower losses and subsequently translates into a higher SNR.

## 2.3. On-bench measurements

On-bench measurements were performed on the constructed metamaterial in order to characterize its frequency response (detailed fabrication process and the geometric information of the metamaterial are provided in the experimental section, Section S1, and Table S1 of the Supporting Information). A vector network analyzer (VNA, P5020B Keysight Inc.) coupled to an inductive pickup loop was employed to characterize the resonance response of the metamaterial. The resonance modes of the metamaterial alone, the DLRR alone, and the combined structure were individually measured under three different excitation strengths, as depicted in **Figure 3**a,b,c, respectively. At an excitation power of -20 dBm, the resonance frequencies of the metamaterial and DLRR alone were tuned to 139.7 MHz and 146.8 MHz, respectively, both higher than the Larmor frequency. This frequency deviation is primarily determined by the strong inductive coupling between the metamaterial and the DLRR, which gives rise to a substantial positive mutual inductance, yielding a collective co-rotating mode with a lower frequency. This coupling was further investigated using the coupled mode theory[37] as detailed in Section S2 of the Supporting Information. As the excitation power gradually increased, the resonance mode of the metamaterial remained unchanged, while the resonance frequency of the DLRR shifted to the left and exhibited a decrease in resonance strength. This frequency shift is a result of the rectifying effect[49,50] in the PIN diodes, which can be viewed as connected in parallel with the distributed capacitance between the conductors. With higher excitation power, the rectifying effect in the diodes acts as a driving voltage, increasing their capacitance, subsequently adding to the self-capacitance of the DLRR, and decreasing the resonance frequency. The decrease in resonance strength of the DLRR is attributed to a strong hysteresis and bistable behavior initially observed in a nonlinear spilt-ring resonator.[49] Nonetheless, the reduced resonance strength in the DLRR is beneficial for minimizing the interaction with the transmission signal for passive detuning purposes.

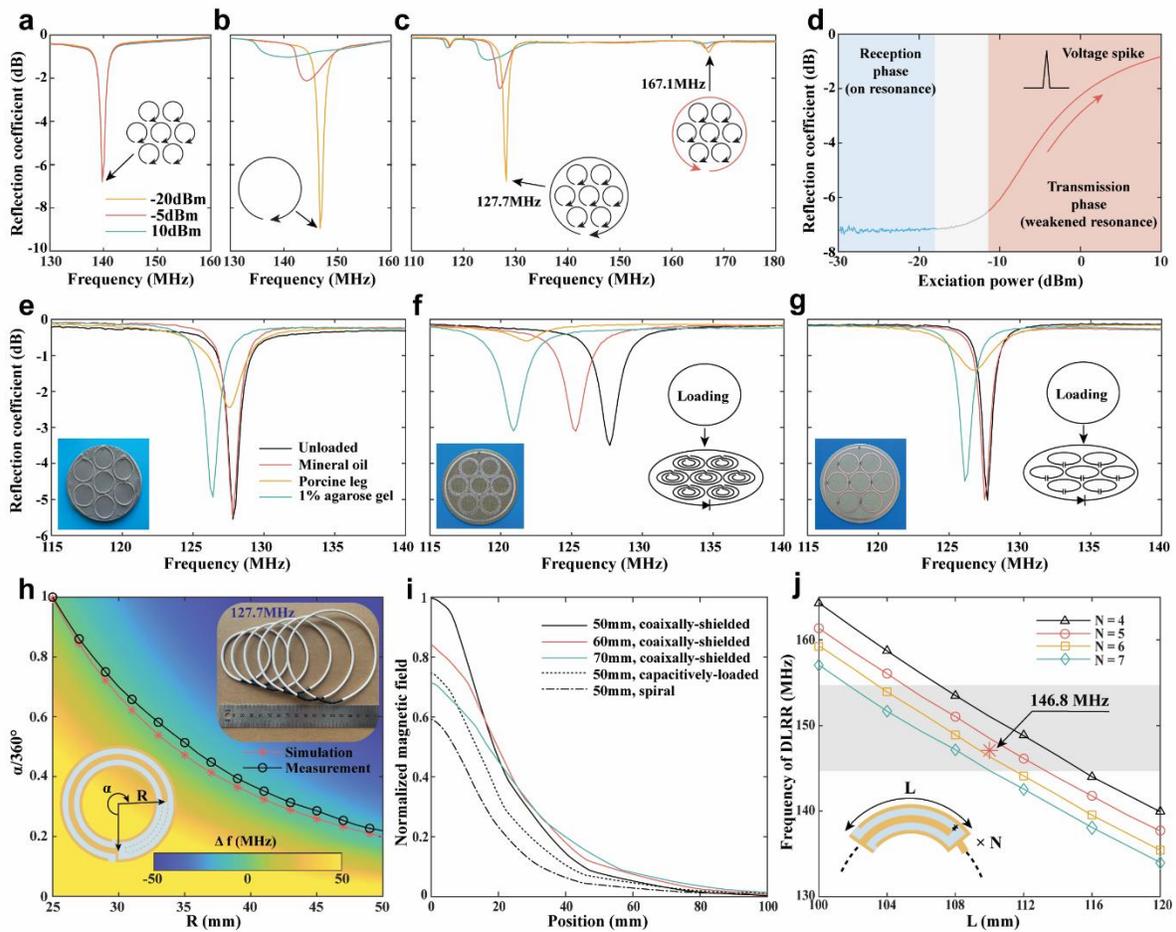

**Figure 3.** Electromagnetic characterization of the metamaterial. a) Measured frequency response of the metamaterial alone, b) the DLRR alone and c) the combined structure under different excitation strengths. d) Measured reflection coefficient as a function of excitation strength at 127.7 MHz. e) Measured reflection coefficient of the coaxially-shielded metamaterial, (f) the spiral metamaterial and g) the capacitively-loaded metamaterial under different loadings. h) Resonance

frequency shift of a unit cell from 127.7 MHz when adopting different radius ($R$) and inner-to-outer conductor length ratios ($\alpha/360°$). The solid lines indicate the simulation and measurement results of parameter combinations that maintain the resonance frequency at 127.7 MHz. Insets: cross-sectional view of a unit cell revealing the frequency tuning mechanism and a series of unit cells adopting different dimensions while all resonating at 127.7 MHz. i) Simulated magnetic field along the center of different unit cells. j) Resonance frequency of the DLRR when adopting different segment numbers (N) and lengths (L). Inset: cross-sectional view of a segment in the DLRR.

When the metamaterial array and the DLRR are placed in close proximity, a strong inductive coupling between them leads to an over-coupled state, resulting in mode splitting, as shown in Figure 3c. Of particular interest is the co-rotating mode (left dip at 127.7 MHz), where the current in the metamaterial and the DLRR flows in the same direction, resulting in an additive magnetic field over the region above the metamaterial. Conversely, the counter-rotating mode (right dip at 167.1 MHz) exhibits destructive interference, yielding a much weaker resonance. Importantly, the close proximity of the components allows the rectifying effect in the PIN diodes to modulate the resonance of the combined structure through inductive coupling (Figure 3c). This was confirmed by sweeping the excitation power of the VNA from -20 dBm to 10 dBm at 127.7 MHz for the combined structure (Figure 3d). At lower excitation levels (< -5 dBm), corresponding to relatively weaker signal strengths (on the order of μW)[51] from the relaxation of precessing protons in the reception phase, the metamaterial's resonance strength remained robust and had the potential to enhance the signal. As the excitation gradually increased, the resonance strength attenuated and exhibited a tendency to eventually stop resonating, thereby remaining silent when exposed to the much stronger signal strength (on the order of kW, typically in the form of transient voltage spikes) generated by RF coils during the transmission phase. Limited by the achievable excitation strength of the VNA, the RF transmitting signal was not able to be fully replicated; the effectiveness of the passive detuning mechanism will be further validated through MRI experiments.

The resonance behavior of the metamaterial was subsequently characterized under various loading conditions to demonstrate its robustness against frequency detuning. The frequency response was recorded with four distinct loadings: a mineral oil phantom, an ex vivo porcine leg sample, a 1% agarose gel phantom, and air (no loading), as shown in Figure 3e. The metamaterial exhibited a minimal frequency shift of 0, 0.2 and 1.4 MHz, respectively, when loaded with the mineral oil, porcine leg, and agarose gel phantom. In comparison, similar measurements were conducted on a planar spiral metamaterial, fabricated using PCB milling techniques on an FR-4 substrate (see section S3 and Figure S2 of the Supporting Information for details). This spiral-based metamaterial closely resembled previously proposed copper conductor-based metamaterials used in MRI,[24-29] lacking inherent geometric properties to effectively confine the electric field. In sharp contrast, it demonstrated a significant susceptibility to frequency shifts, with the mineral oil, porcine leg, and agarose gel phantom loadings resulting in shifts of 2.4, 5.8, and 6.8 MHz, respectively (Figure 3f). An alternative approach to address the frequency detuning challenge involves the use of a capacitively-loaded single-turn loop-based metamaterial design,[52] capable of confining nearly all of the electric field at the capacitor, effectively lowering the metamaterial's operating frequency (details in section S3 of the Supporting Information). The frequency response under different loadings of this capacitively-loaded metamaterial is depicted in Figure 3g, exhibiting performance similar to the coaxially-shielded metamaterial, with frequency shifts of 0.2, 1 and 1.6 MHz when loaded by the three phantoms. However, the capacitively-loaded metamaterial design presents several drawbacks compared to the proposed coaxially-shielded metamaterial. These include reduced resilience against deformation, the need for delicate handling, increased cost, intricate fabrication, limited choices for low-loss and flexible substrates, and discrete achievable metamaterial dimensions due to discrete capacitor values. Most importantly, however, the capacitors contribute to resistive losses within the metamaterial, resulting in a lower SNR enhancement, as will be demonstrated in the MRI validations.

## 2.4. Frequency tuning mechanism

Matching the resonance frequency to the Larmor frequency is essential for the operation of the metamaterial as a frequency mismatch can result in suboptimal performance or even artifacts.[26] Additionally, to achieve a snug fit about a curved anatomical region, such as the human knee, the metamaterial must be bent into a curved configuration, as illustrated in Figure 1c. In this arrangement, it is inevitable that the magnetic flux in neighboring unit cells will overlap, potentially causing a slight shift in the resonance frequency (refer to Section S4 and Figure S3 of the Supporting Information). Therefore, it becomes crucial to consider a frequency tuning mechanism to compensate for such shifts and thereby broaden the metamaterial's applicability across various anatomical regions.

The construction of the metamaterial reported herein using coaxial cable has eliminated the need for the use of a lumped capacitor because a structural capacitance naturally forms between the inner and outer conductor. Therefore, it is possible to tailor the length of the inner conductor (inner-to-outer conductor length ratio) to fine tune the resonance frequency of the metamaterial.[53] In the proposed design, the unit cell has an inner-to-outer conductor length ratio of 1 and a radius of 25 mm, resulting in a unit cell frequency of 123 MHz. By placing 7 such unit cells in close proximity, the inter-unit cell coupling gives rise to a negative mutual inductance and mutual capacitance, resulting in a collective mode with a higher resonance frequency at 139.7 MHz (Please see Section S2 of the Supporting Information for details). Figure 3h illustrates how the resonance frequency of a single metamaterial unit cell can be shifted by varying the radius and the inner-to-outer conductor length ratio. With a fixed resonator radius, precise control over the resonance frequency can be achieved by adjusting the length of the inner conductor. Additionally, any frequency shifts resulting from bending, stretching, coupling to the BC, loading, etc., may be conveniently compensated for. Moreover, this tuning mechanism provides flexibility to design metamaterials with specific dimensions tailored to particular requirements (Experiments conducted using a metamaterial with broader coverage can be found in Figure S4 of the Supporting Information). In contrast, capacitively-loaded metamaterials can only achieve discrete dimensions due to the discrete values of capacitors. The solid lines in Figure 3h represent parameter combinations that maintain the resonance frequency at 127.7 MHz. Furthermore, Figure 3h also displays a series of coaxially-shielded loop resonators fabricated using the proposed technique, all sharing the same resonance frequency of 127.7 MHz but featuring different dimensions. Figure 3i demonstrates the simulated normalized magnetic field enhancement ratio at the center of three coaxially-shielded loop resonators with diameters of 50, 60, 70 mm, respectively. It is evident that by adjusting the unit cell size, a balance can be struck between magnetic field enhancement, penetration depth, and spatial coverage, making it adaptable to various MRI application scenarios (Please see Section S5 and Figure S5 of the Supporting Information for further details). In addition, the magnetic field enhancement capability of a coaxially-shielded resonator is compared to a capacitively-loaded resonator and a spiral resonator, both having a diameter of 50 mm. The coaxially-shielded resonator exhibits a higher degree of magnetic field enhancement owing to its low intrinsic loss and high quality factor.

Tuning the working mode of the metamaterial to the Larmor frequency of the 3.0 T MRI is achieved through coupling when positioning the metamaterial and the DLRR in close proximity. While there are no strict limitations on the frequency of the DLRR to allow passive detuning for the combined structure, it is necessary to also introduce a tuning mechanism for the DLRR design to adapt to various application scenarios. The DLRR consists of five segments of diode-loaded coaxially-shielded resonators, each with a length of 110 mm, resulting in a ring resonator with a resonance frequency of 146.8 MHz and a diameter of 175 mm. This size is comparable to the metamaterial array (165 mm), thereby promoting strong inductive coupling when placed in close proximity to the metamaterial and enabling a frequency match to 127.7 MHz for the combined structure. However, both the number of segments ($N$) and their length ($L$) can vary to accommodate alternative array dimensions. Figure 3j depicts the simulated relationship between $L$ and the resonance frequency of the DLRR while adopting 4, 5, 6, and 7 segments in the DLRR, demonstrating an almost linear relationship. In practice, we have determined that tuning the DLRR to a frequency range of 145-155 MHz is adequate to match the working mode of the metamaterial to the Larmor frequency of 3.0 T MRI. It is worth noting that changes in both $N$ and $L$ not only directly affect the resonance frequency but also influence the coupling strength between the DLRR and the metamaterial in the combined structure. Therefore, customizing the DLRR design for specific cases is advisable to ensure optimal performance and resonance matching.

## 2.5. Validations of the metamaterial in simulation

To characterize the performance of the metamaterial in an MRI system, a high-pass birdcage coil (BC) was built in simulation, operating as both the transmitter and the receiver. The BC was tuned to 127.7 MHz when the metamaterial is positioned at its isocenter. In simulation, it is challenging to replicate the excitation power-dependent frequency response of the PIN diode, along with its hysteresis and bi-stable response. Therefore, we replaced the PIN diodes with capacitors, modifying their capacitance to achieve the desired frequency shift, thus determining the optimal conditions for reception and transmission phases. The metamaterial was applied in a curved configuration, wrapping around a mineral oil phantom with a 130mm diameter. The simulated magnetic field in the transverse and coronal planes are depicted in **Figure 4**a,b. The metamaterial exhibited a remarkable magnetic field enhancement of up to 110 times within the region of interest (ROI) for the receiving field $B_1^-$. In contrast, the transmitting field $B_1^+$ remained almost unchanged. It is important to note that while the SNR enhancement hinges on the magnitude of the secondary magnetic field generated by the metamaterial, the field direction does not affect the signal acquisition. Since both $B_1^-$ and $B_1^+$ are circularly polarized fields rotating in the transverse plane, the metamaterial remains effective as long as it is not primarily aligned in the transverse plane. Additionally, the phase information of the magnetic field along the dotted arrow is also plotted in Figure 4a, indicating no phase transition induced by magnetic field cancelation inside the ROI, further validating that the magnetic field of the metamaterial and the DLRR interfere constructively for the co-rotating mode. Furthermore, a human voxel model is adopted to assess the safety of the proposed metamaterial. The SAR map (10g) was calculated in proximity to the knee joint and depicted in Figure 4c,d, indicating that the metamaterial is as safe as using the BC alone. This is primarily attributed to the passive frequency detuning and the strong electric field confinement of the metamaterial.

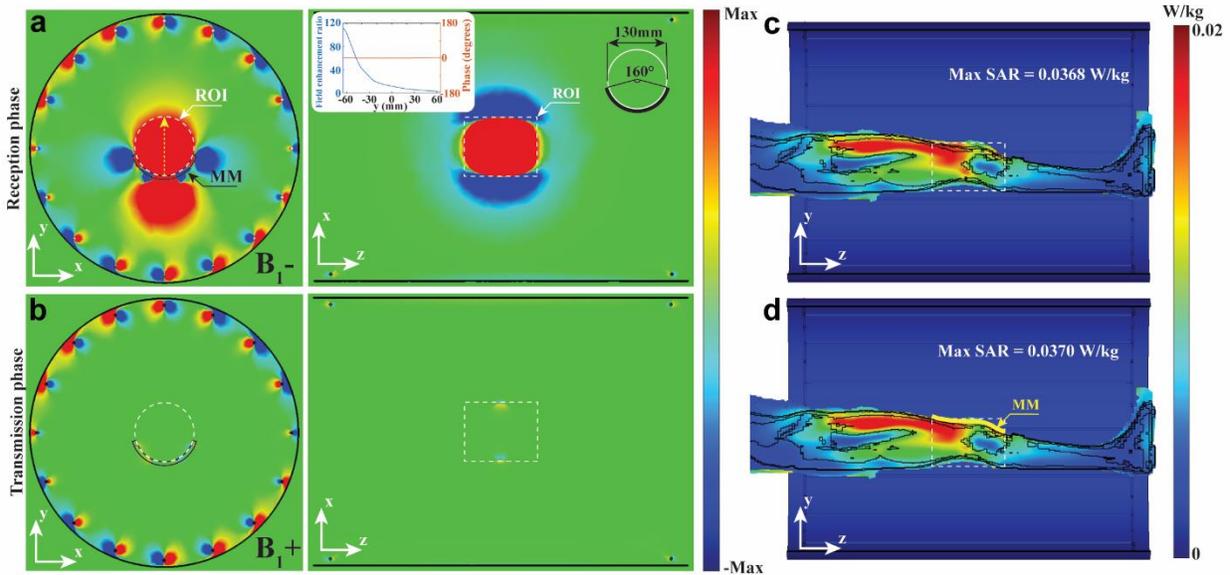

**Figure 4.** Characterization of the metamaterial in simulation. a) Simulated magnetic field distribution in the transverse and coronal planes during the reception phase and b) the transmission phase. Inset: magnetic field enhancement and phase information along the dashed arrow. c) Simulated SAR maps (10g, normalized to 1W of accepted power) acquired with the BC alone and d) with the BC and the metamaterial in the sagittal plane.

## 3. Experimental MRI validation

### 3.1. MRI validation with a mineral oil phantom

MRI validations were firstly performed on the metamaterials in a flat configuration using a clinical 3.0 T MRI scanner (Philips Healthcare). A mineral oil phantom with a diameter of 150 mm was placed directly above the metamaterial and moved to the isocenter of the BC. When the metamaterial is used for imaging, the BC was used as both the transmitter and the receiver. In addition, a reference case in which the phantom was imaged using a commercially available single-channel surface coil (Flex M coil, Philips Healthcare) was used for comparison; the single channel coil has a coverage of 150 mm, which is close to that of the metamaterial (175 mm). The SNR maps were depicted in **Figure 5**a,b, in which the dotted lines indicate the paths along which the SNR was evaluated. The SNR profiles were compared at three positions with the SNR values depicted in Figure 5c and normalized to the case in which the phantom was imaged using the BC alone. At the center of the metamaterial, the peak SNR enhancement ratio was approximately 2.5-fold when compared to the single-channel surface coil, and 42-fold compared to the BC alone. Notably, the metamaterial consistently outperformed the single-channel surface coil throughout the entire phantom, with a 36% and 33% enhancement at 50 mm and 100 mm inside the phantom, respectively. In the region between the neighboring metamaterial unit cells, a phase transition and a spatial 'dark spot' inevitably occurred, resulting in a 25% lower peak SNR than the flex coil. Nevertheless, this SNR reduction was effectively compensated for by the magnetic field generated by the DLRR (Please see Section S6 and Figure S6 of the Supporting Information for further details), and extended only 10 mm into the phantom; the metamaterial's performance matched that of the Flex M coil with a 20% enhancement beyond this region. Even at the edge of the phantom, where the SNR of the single-channel coil was twice as high as that at the center due to its close proximity to the copper trace, the single-channel coil was still outperformed by the metamaterial. In addition, the analytical results of the SNR enhancement ratio were also derived from simulation and plotted in Figure 5c, demonstrating a high degree of agreement with the experimental results (the derivation of the analytical SNR values is explained in detail in the experimental section). Subsequently, the performance of the coaxially-shielded metamaterial is also compared against two alternative designs: the capacitively-loaded metamaterial and the spiral metamaterial. A 20% and 64% higher peak SNR was observed for the coaxially-shielded metamaterial when compared to these metamaterials, respectively, as illustrated in Figure 5d and Figure S2 of the Supporting Information. This enhanced performance can be primarily attributed to the lower dielectric loss, higher quality factor, and the absence of additional loss from lumped elements in the coaxially-shielded metamaterial design.

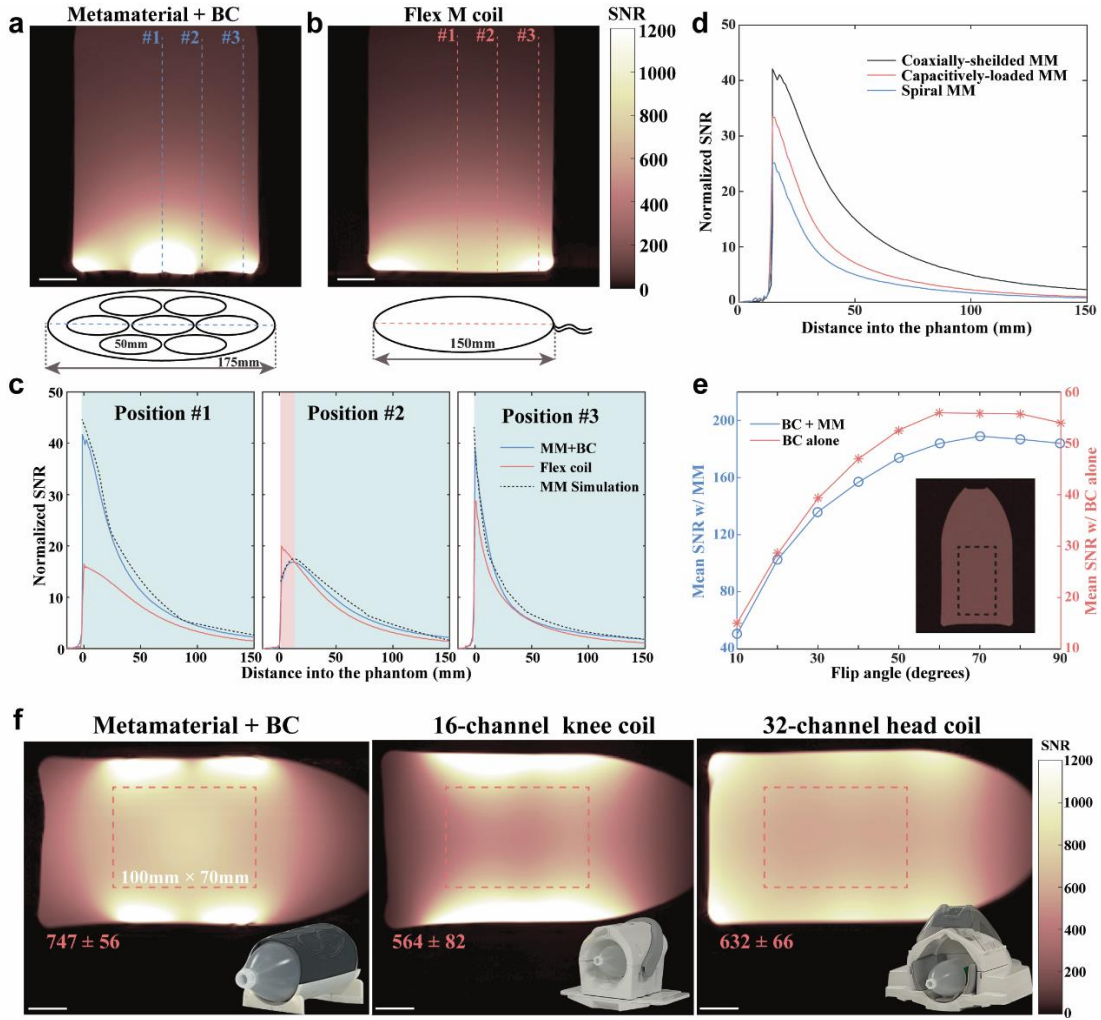

**Figure 5.** MRI validations using a mineral oil phantom. a) Axial SNR map acquired using the BC with the metamaterial and b) using the single-channel receive coil. c) SNR profiles along the dashed lines in (a) and (b). d) SNR profiles along the center of the phantom acquired with three different metamaterial configurations. e) Mean SNR (evaluated in the dashed rectangular region) as a function of the FA with and without the metamaterial. f) Coronal SNR maps acquired with the metamaterial, the 16-channel knee coil and 32-channel head coil. The SNR (average ± standard

deviation) is evaluated in the outlined dashed region. Insets: experimental setup for the mineral oil phantom imaging. Scale bars in (a), (b) and (f) are 3 cm.

When the metamaterial was employed during MRI scans, its interaction with the transmitting field B1+ may be reflected by the resultant flip angle (FA) in the phantom, as the FA scales linearly with $B_1^+$.[54] Ideally, since the metamaterial is passively detuned and remains in an off-state during the transmission phase, the FA should remain unaffected in the presence of the metamaterial. The mean SNR within the dashed area in Figure 5e was evaluated by varying the FA. In both cases, with and without metamaterial, while the SNR variation is significant under different FAs, they exhibit the same trend, further confirming a negligible interaction between the metamaterial and $B_1^+$.

The performance of the metamaterial was further validated in a curved configuration by wrapping and conforming it to a cylindrical mineral oil bottle phantom with a diameter of 130 mm, demonstrating its applicability in clinical knee imaging. The SNR map of the coronal plane when using the metamaterial is depicted in Figure 5f. To put the performance of the coaxially-shielded metamaterial into perspective, it was compared against two state-of-the-art multi-channel coils: a 16-channel transceiver knee coil and a 32-channel receiver head coil. The SNR maps of the same coronal plane using these commercial coils are also shown in Figure 5f. The SNR was then evaluated in the dashed rectangular region (100 mm × 70 mm), where the metamaterial demonstrated a 32% higher SNR than the knee coil and an 18% higher SNR than the head coil, while exhibiting a more homogeneous signal distribution. The performance of the metamaterial was also evaluated in the sagittal and axial planes, as shown in Figure S7 of the Supporting Information; while a high SNR is maintained, inevitably, an SNR gradient is present due to the arrangement of the metamaterial, yielding a less uniform SNR distribution than the head and knee coils. However, the metamaterial's wireless and flexible nature imposes no restrictions on patient positioning in practice. Therefore, it becomes feasible to obtain high and homogeneous SNR for imaging various planes of the knee or other anatomy of interest, thus offering versatility in clinical applications.

### 3.2. MRI validation with a porcine leg

Finally, an ex vivo porcine leg sample was adopted to test the metamaterial in a more biomedically relevant environment. We adopted different pulse sequences commonly used in day-to-day routine clinical imaging to demonstrate the metamaterial's sequence compatibility enabled by the DLRR. The same porcine leg was scanned with the metamaterial and the 16-channel knee coil using 5 common clinical pulse sequences including gradient echo (GRE), spin echo (SE), T1-weighted turbo spin echo (T1w TSE), T2-weighted turbo spin echo (T2w TSE) and proton density-weighted turbo spin echo (PDw TSE) imaging. MRI images of these scan results are depicted in **Figure 6**a while the experimental setup is depicted in Figure S8 of the Supporting Information.

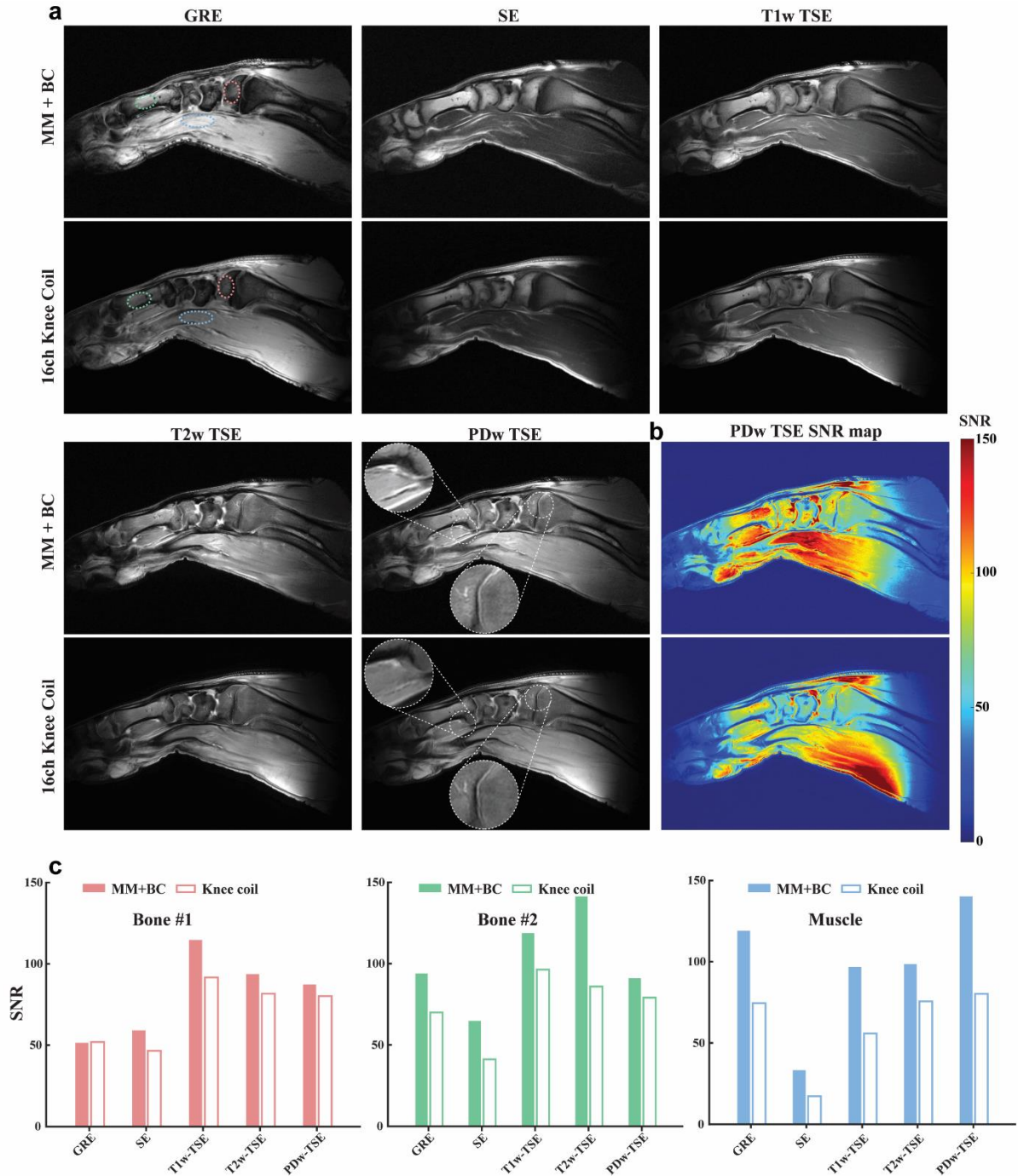

**Figure 6.** MRI validations with the ex vivo porcine leg sample. a) MRI images of the porcine leg acquired with the metamaterial and the 16-channel knee coil using 5 MRI pulse sequences: GRE, SE, T1w TSE, T2w TSE and PDw TSE. b) SNR map of the PDw TSE images. c) Quantitative evaluation of the SNR for different regions as outlined in (a).

The results indicate that the metamaterial is compatible with all the pulse sequences. When compared to the MRI images acquired by the knee coil, the metamaterial is able to produce image quality that is comparable or even superior in some aspects. For example, in the PDw TSE image acquired with the metamaterial, detailed features including the cartilage, intermuscular fascia and the periosteum are visualized remarkably well, surpassing the image quality obtained with the knee coil. Importantly, the metamaterial does not introduce any noticeable artifacts. Furthermore, the metamaterial offers a broader field of view since the signal is received by the BC with a broader coverage than the knee coil. The SNR maps of the PDw TSE images were then calculated and plotted in Figure 6b, noting that the metamaterial promotes a homogeneous SNR enhancement across the entire knee joint. In contrast, the knee coil primarily exhibits significant SNR enhancement in the peripheral regions, mainly composed of skin, muscle and fat. To provide a quantitative perspective, we evaluated the SNR in three distinct regions, including two bone segments and one muscle tissue segment, as indicated by the dashed circles in Figure 6a. The quantitative SNR values demonstrate that, across different sequences, images acquired with the metamaterial show an SNR improvement of up to 2-fold in these regions when compared to the knee coil. It is worth noting that the 16-channel knee coil used in the experiment, while designed with ergonomic considerations for various anatomical sizes, inevitably trades off sensitivity for applicability due to its fixed and rigid design. In contrast, the metamaterial has the potential to maximize sensitivity by conforming closely to the anatomy, delivering superior performance even when operating wirelessly. Most importantly, this advantage is attributed to the inherited strong field enhancement of individual unit cells and the expanded coverage achieved through the co-rotating collective resonance mode. Furthermore, the metamaterial weighs only 50 g, including the housing fabric, ensuring a comfortable clinical scan experience regardless of the imaging configuration. Additional MRI scans conducted on a smaller biological sample, a chicken drumstick, are detailed in Section S7 and Figure S9 of the Supporting Information, demonstrating the capability of the metamaterial for imaging a smaller anatomic region.

## 4. Discussion

In this study, for the first time, we developed a wearable magnetic metamaterial crafted from commercially available coaxial cable and demonstrated its exceptional SNR enhancement capabilities within a 3.0 T MRI system. By leveraging the distributed capacitance formed between the two conductor layers of the coaxial cable, the fabricated metamaterial can operate seamlessly at the Larmor frequency of 3.0 T MRI without the need for any lumped elements. Meanwhile, the metamaterial effectively confines its electric field within the coaxial cable layers, mitigating the undesired electric coupling to the sample. This electric field confinement was then translated into a lower loss and a marked SNR enhancement.

Notably, our previously proposed helix coil-based metamaterials[24-26] also exhibit a SNR enhancement capability but MRI validations were less diagnostically relevant, as the comparison was made to the BC, which has significantly lower sensitivity compared to state-of-the-art receive coil arrays typically used in clinical imaging. While some alternative metamaterial forms also exhibit the potential for achieving SNR comparable to clinically available receive coil arrays,[29,30] their SNR enhancement capacity hinges on the suboptimal application of the reference coil and the very compact arrangement of the rigid and bulky unit cells. Consequently, their applicability is restricted to either small anatomical areas such as the human wrist or lower magnetic field MRI systems, typically 1.5 T. During the development of the coaxially-shielded metamaterial, these limitations were actively addressed. The metamaterial's performance was first validated using a mineral oil phantom in both flat and curved configurations, demonstrating its broad applicability spanning from extensive, flat anatomic regions such as the chest, abdomen and pelvis to more confined small, curved regions such as the knee, ankle or wrist. Additionally, the proposed metamaterial extends its applicability in both 1.5 T and 7.0 T MRI systems (see Section S8 and Figure S10 of the Supporting Information). Through scans of an ex vivo porcine leg sample, the performance of the metamaterial was compared to a state-of-the-art 16-channel transceiver knee coil, resulting in comparable images and a notable SNR enhancement of up to 2-fold. Most importantly, the incorporation of the DLRR, inductively coupled to the metamaterial, enables detuning of its resonance mode during the transmission phase, allowing the metamaterial to be seamlessly integrated into day-to-day clinical imaging.

The proposed metamaterial consists of 7 unit cells arranged in a hexagonal configuration, the DLRR adopts a circular design to closely match the contour of the metamaterial. Nevertheless, we should emphasize that the shape, number, and arrangement of these unit cells are highly adaptable and should be customized to meet specific application requirements. The hexagonal configuration presented herein was selected based on practical testing, which demonstrated that it results in densely packed unit cells, leading to a more concentrated magnetic field and a higher SNR. For reference, we also constructed an alternative design consisting of 9 square unit cells with a square-shaped DLRR comprising 6 segments. This alternative configuration is capable of achieving a comparable SNR to the proposed setup (Figure S11 of the Supporting Information). This flexibility in design highlights the versatility and adaptability of the metamaterial to suit various imaging scenarios and requirements.

The introduction of passive detuning capability in the proposed metamaterial is crucial because it ensures compatibility with commonly used clinical imaging sequences. While previous metamaterial designs lacking this detuning feature have been proposed, along with manual calibration techniques for adjusting transmitting power to make them usable,[27,28] their optimal performance is generally limited to one specific imaging plane parallel to the metamaterial. Moreover, in clinical practice, determining and maintaining the optimal transmitting power manually can be highly unrealistic and impractical due to the dynamic and variable nature of patient anatomy and positioning. In our deign, we have achieved passive detuning by the usage of a DLRR that operates via inductive coupling with the metamaterial, leveraging the rectifying effect and the resulting excitation-dependent frequency response of the PIN diodes. Importantly, the DLRR itself, acting as a magnetic ring resonator, also contributes to the total SNR enhancement, especially for further distances. An alternative approach to achieve passive detuning involves directly integrating the PIN diode inside the metamaterial unit cell. However, this integration introduces a series resistance into the metamaterial, leading to a lower SNR. Additionally, in the absence of the DLRR, the penetration depth of the metamaterial is significantly reduced, resulting in a much lower SNR for objects located at greater distances. Please see Section S6 and Figure S6 of the Supporting Information for further details.

## 5. Conclusion

The proposed coaxially-shielded metamaterial offers a novel paradigm for the construction of magnetic metamaterials for MRI, paving the way for the development of a next-generation wireless MRI technology. Future improvement of the proposed work involves ergonomic considerations for targeted MRI, enabling a form-fitting design that can adapt to an arbitrarily shaped anatomical shape. This approach aims to provide a wireless, lightweight, cost-effective, fast and comfortable MRI solution, potentially enabling studies of joint motion during flexion or real-time MRI applications. Furthermore, the operational frequency of the metamaterial, primarily determined by the geometric dimensions of the commercially available coaxial cable groups, falls within the MHz range, conveniently facilitating its integration into MRI systems. Consequently, some of its inherently unique characteristics, including the resilience against frequency detuning, and reduced loss due to the elimination of eddy currents, offers promising pathway for the development of electromagnetic devices in other scenarios involving radio frequency near-field powering or communication.

### *Experimental section:*

***Metamaterial construction***: The metamaterial unit cell was fabricated using a 170 mm-long segment of coaxial cable. At one end, a small portion of the inner conductor is removed to prevent contact with the outer conductor, while the other end had a portion of the outer conductor removed to expose the inner conductor. Subsequently, the exposed inner conductor was then welded to the outer conductor on the opposing end, with reinforcement through heat shrink tubing. The resonance frequency of an isolated unit cell is 123 MHz, however, when the 7 unit cells are placed in proximity, the strong inductive coupling gave rise to a substantial negative mutual inductance and mutual capacitance, resulting in a collective resonance mode with a higher frequency of 139.7 MHz. The construction of the DLRR is similar to that of the metamaterial except that a PIN diode was welded at the gap between the inner and outer conductor in each segment.

The employed non-magnetic coaxial cable belongs to the RG174 cable group. The diameter of the inner conductor, the insulating dielectric, the outer conductor and the jacket are 0.48, 1.42, 1.93 and 2.54 mm, respectively. The selection of this cable type was made considering factors such as cable flexibility, resistance, and ease of fabrication. While other commonly encountered cable groups like RG178 and RG58 also exhibit potential for use in metamaterial construction, their suitability involves a tradeoff between various properties. Further details can be found in Section S8 and Figure S11 of the Supporting Information.

The bonding of the two layers of the housing fabric was accomplished by utilizing a digital embroidery machine (PE535, Brother) to create a pattern that followed the contour of the metamaterial, ensuring a seamless integration of the metamaterial within the fabric layers. The capacitively-loaded metamaterial and spiral metamaterial were fabricated on the FR-4 substrate with a PCB prototype machine (ProtoMat S64, LPKF).

***Numerical simulation***: All electromagnetic simulations related to the magnetic field distribution were performed using the frequency domain solver in CST Microwave Studio Suite 2021. The SAR mapping was obtained using the time domain solver with the human voxel model 'Gustav' form the CST voxel family. A high-pass BC was built to provide a circularly polarized field; the BC was comprised of 16 800 mm-long legs, 32 lumped capacitors with a capacitance of 10pF, and 2 discrete ports with a 90° phase shift. The mineral oil phantom was constructed with the following parameters: diameter 120 mm, height 200 mm, relative permittivity 2.1, and conductivity 0.175 S/m.

The simulation results were used to evaluate the optimal SNR enhancement ratio in the phantom. The expression for SNR in MRI is[55]:

$$SNR \propto \frac{\omega^2 B_c}{\sqrt{R_c + R_{MM} + R_p}} \qquad (1)$$

in which $\omega$ is the Larmor frequency of the MRI and $B_c$ is magnetic field generated with unit current in the BC when it is working in a receiving mode. $R_c$ is the resistance of the receiving coil while $R_{MM}$ and $R_p$ are the equivalent resistance induced by the metamaterial and the phantom in the receive coil. The power dissipation from these three resistance terms is directly proportional to each respective resistance. Therefore, the SNR can be expressed as:

$$SNR \propto \frac{\omega^2 B_c}{\sqrt{P_c + P_{MM} + P_p}} \qquad (2)$$

where $P_c$, $P_{MM}$ and $P_p$ represent the equivalent power dissipation induced in the receive coil by the receive coil itself, the metamaterial and the phantom. $B_c$ was determined by comparing simulations with and without the metamaterial, and power dissipation values are extracted from the CST 1D results as material-specific power loss.

In terms of the SAR, it was derived using the time domain solver with and without the metamaterial. The SAR represents the rate of power dissipated in certain mass of tissue induced by an external RF excitation, which is expressed by[48]:

$$SAR = \frac{\sigma E^2}{2\rho} \tag{3}$$

in which $\sigma$ and $\rho$ are the conductivity and density of the tissue. $E$ represents the peak amplitude of the electric field induced by the transmitting RF field. The SAR was normalized to 1W of accepted power during the simulation.

***MRI validation***: All MRI experiments performed using the mineral oil phantom adopted the gradient echo pulse sequence for imaging. The FA was set to 90° for all phantom experiments unless otherwise indicated. Detailed information about the experimental setup and sequence parameters of the MRI validations can be found in Section S9 and Table S2 of the Supporting Information.

## Supporting Information

Supporting Information is available from the Wiley Online Library or from the author.

## Acknowledgments

Xia Zhu and Ke Wu contributed equally to this work. This research was supported by the Rajen Kilachand Fund for Integrated Life Science and Engineering. The authors thank Dr. Yansong Zhao for technical assistance and valuable discussions during MRI experiments. The authors thank Boston University Photonics Center for technical support.

## Conflict of interest

The authors have filed a patent application on the work described herein, application No.: 16/002,458 and 16/443,126. Applicant: Trustees of Boston University. Inventors: Xin Zhang, Stephan Anderson, Guangwu Duan, and Xiaoguang Zhao. Status: Active.